\documentclass[%
superscriptaddress,
amsmath,amssymb,
reprint,
nofootinbib,
]{revtex4-2}

\usepackage{caption}
\usepackage{subcaption}
\usepackage{hyperref}
\usepackage{siunitx}
\usepackage{cleveref}
\usepackage{dcolumn}

\usepackage{hyperref}
\hypersetup{
  colorlinks=true,
  citecolor=blue,
  linkcolor=blue,
  urlcolor=blue
}

\usepackage[utf8]{inputenc}
\usepackage{graphicx}
\usepackage{enumitem}
\usepackage{xcolor}
\usepackage{dsfont}
\usepackage{listings}

\newcommand{\crayfis}{{\sc Crayfis}}

\begin{document}


\title{ Data Acquisition System for a Distributed Smartphone Cosmic Ray Observatory }
\author{Jeff Swaney}
\affiliation{Department of Physics and Astronomy, University of California, Irvine, CA 92697, USA}
\author{Chase Shimmin}
\affiliation{Yale University, New Haven, CT USA}
\author{Daniel Whiteson}
\affiliation{Department of Physics and Astronomy, University of California, Irvine, CA 92697, USA}

\begin{abstract}
A scientific instrument comprised of a global network of millions of independent, connected, remote devices presents unique data acquisition challenges. We describe the software design of a mobile application which collects data from smartphone cameras without overburdening the phone's CPU or battery. The deployed software automatically calibrates to heterogeneous hardware targets to improve the quality and manage the rate of data transfer, and connects to a cloud-based data acquisition system which can manage and refine the operation of the network.
\end{abstract}

\maketitle

\tableofcontents

\section{Introduction}

The rise of smartphones has transformed the nature of daily life, creating an affordable, portable networked computing platform, which has been adopted by a significant fraction of the Earth's population~\cite{phoneusage}.  
If it could be repurposed as a scientific instrument, this enormous collective investment in distributed data collection devices could transform some areas of cutting-edge research.  
While projects have previously made use of the untapped computing power of consumer devices~\cite{setiathome,Amorim:2004cz,parasitic-computing} or enlisted users for crowdsourced observational studies~\cite{ebird,socialwater}, a further frontier exists in using the devices to actually {\it create} new data through the use of the on-device instruments. 
Such a network represents an enormous resource, whose cost is comparable to the entire global scientific research budget, but also presents novel challenges for scientific applications, which often have stringent operational requirements.

Contrary to instruments and data acquisition systems in most scientific settings, the global network of smartphones presents a heterogeneous platform, comprised of hundreds of hardware models from dozens of manufacturers running many different versions of software.  
A second challenge is that the experimenter will likely never have direct any physical access to the devices, requiring automatic remote calibration and operation. 
Any scientific application would operate as an adjunct to, rather than the primary purpose of, the existing smartphone network. 
Therefore the experiment's use as a data collection platform must be made unobtrusive for the device owners by limiting interference with normal operations, excess network utilization, and adverse effects on device hardware and lifespan.

The possible scientific applications of such a network are vast, and the repurposing of on-device instruments for scientific measurements has only recently begun to be explored. 
Examples include using the CMOS camera as a particle detector~\cite{crayfis2015} or radiation sensor~\cite{Cogliati:2014uua,KANG2016126}, the accelerometer as a seismometer \cite{myshake}, or the barometer for weather forecasting \cite{pressurenet}. 
Tying together a network of phones into a single large scientific instrument requires lightweight, robust, self-calibrating on-device software combined with a global, flexible data acquisition system which can collect and manage the data. 
It is also important to monitor devices to provide feedback to optimize per-device performance as well as to coordinate and regulate activity network-wide.  
In this paper, we describe the design and implementation of such a system for \crayfis~\cite{crayfisurl}, an experiment which seeks to observe extended air showers created by high-energy cosmic rays.

\section{Experimental Requirements}

Ultra-high energy cosmic rays (UHECRs) are particles incident upon Earth's atmosphere with energy above \SI{e18}{eV}.  
These rare events present an enduring scientific mystery. 
Despite many theoretical conjectures~\cite{Bell:1978zc,Blandford:1987pw,Waxman:1995vg,Weiler:1997sh}, their astrophysical source remains unexplained~\cite{Abraham:2007si,Abraham:2007bb}.  
When the energetic primary particle of a UHECR interacts with molecules in the atmosphere, great numbers of lower-energy particles are produced, forming extensive air-showers.  
These cascades of secondary particles are detectable via their particle flux on the ground, fluorescence in the air, or radio and acoustic signatures.

Dedicated facilities~\cite{Abbasi:2002ta,Takeda:1998ps,Abraham:2008ru,Aab:2014ila,AbuZayyad:2012ru,Hayashida:1998qb} observe cosmic
rays, reporting energies up to  $3\cdot
10^{20}$ eV. 
Such high-energy observations are rare and precious due to interaction with the cosmic microwave
background~\cite{Greisen:1966jv,Zatsepin:1966jv} which suppresses particle flux precipitously above $5\cdot 10^{19}$ GeV. 
Accumulating larger samples of UHECRs using traditional mechanisms would require significantly longer observations or larger facilities.

The CMOS sensors underlying smartphone cameras have been shown to be sensitive to ionizing radiation~\cite{2002SPIE.4669..172S,crayfis-muons,crayfis-sim}, which deposits energy in silicon photodiodes, producing bright isolated pixels or small clusters over an otherwise dark background.
Ref.~\cite{crayfis2015} explored the feasibility of employing the world-wide network of billions of consumer smartphones to detect the passage of air-showers through deposition of energy in the CMOS sensors. 
A worldwide network could complement the observational power of existing facilities, as well as offer sensitivity to coincident events that are correlated on a global scale from novel phenomena~\cite{Albin:2021psb}. 

Surface arrays such as \crayfis~rely principally on muons, photons, and electrons produced in extended air-showers, as the flux of charged hadrons is greatly attenuated with atmospheric depth.
From the efficiencies provided by Refs.~\cite{crayfis-muons,crayfis-sim}, a sufficiently dense array of smartphone cameras will detect significant quantities of each of these species in the event of a UHECR shower. 
However, the CMOS sensors in smartphones are likely too thin to provide useful calorimetric data or to reliably distinguish between these particle species. 
Terrestrial radiation and muons produced by lower-energy cosmic rays therefore constitute an unavoidable combinatorial background which must be distinguished from shower constituents by requiring simultaneous, geographically-clustered detections~\cite{crayfis2015}. 

By measuring the shower density across a local cluster of phones spanning several kilometers, the energy, direction, and mass of the primary particle can be estimated.
The network's power therefore depends on each phone's efficiency to observe individual particles, and ability to suppress false positives due to noise.
Equally vital are non-observations, which can help reject false positives from nearby devices and can be used to set upper bounds on the rate of rare events. 

The most basic requirement for each device is then to continuously monitor the image sensor and provide a stream of data indicating observations and non-observations with location and time information. 
The software should automatically select a {\it trigger threshold} to determine whether a given image frame is classified as an observation, in a way which balances efficiency to identify true particle events and minimizes false positives. 
Crucially, the efficiency of each devices' trigger must be measurable \textit{in situ}.  
To optimize detection efficiency, the full sensor area must be utilized with a trigger correcting for inhomogeneity in the response; for example, persistently noisy ``hot pixels'' of the CMOS, which would otherwise dominate the stream, should be automatically identified and suppressed on-device to conserve network bandwidth and device CPU. 
These on-device activities are subject to the limited computing constraints of an unobtrusive mobile application.

Data captured by each device must be transferred to a central data acquisition (DAQ) service. 
In addition to archiving recorded data, the DAQ system can perform more computationally-intensive analysis of the patterns of pixel rates, to help identify poorly performing pixels and provide feedback to each device to improve its data quality. 
Long latencies between collection and transmission can be tolerated, especially if it minimizes the use of more expensive network connections. 
Long gaps in data uploads may minimize the ability of the DAQ to provide feedback on the data quality, however.

\section{Device configuration and data collection}

The CMOS sensor is comprised of a set of pixels, typically $\sim$\SI{1}{\micro\metre} in pitch and \SIrange[range-phrase=--]{2.5}{3.5}{\micro\metre} in depth~\cite{Fontaine2019}. 
Each pixel contains one or more silicon photodiodes: $p-n$ junctions which convert  deposited energy into mobile electron-hole pairs, which are collected and measured as an accumulated charge. 
Electron-hole pairs can also be generated spontaneously via thermal excitation, creating a dark current. 
The pixels are designed for use in visible-light photography, and so focusing lenslets and a color filter array sit atop each pixel. 
The measured charge is transformed into a digital pixel value through application of a gain, subtraction of black level, and digitization, typically at 10-bit precision. 
In many models, a subset of pixels is dedicated to calibrating auto-focus through partial physical masking~\cite{Fontaine2019} and other such means. 
The sensor controller may report interpolated values for these pixels rather than their true measured photocurrents. 

The digitized pixel values comprise a RAW-format image, which is ideal for professional photographers and particle physicists; however, the image pipeline is not optimized for processing these buffers at high frame rates, and the high resolution imposes significant burdens on the CPU and memory to analyze on-device. 
In more typical video buffering, the frame-processing pipeline converts the single-color-channel pixels of the RAW image into 8-bit RGB values, using various downsampling and interpolation techniques to reduce the resolution and color depth.  
For instance, spatial downsampling may occur via binning, in which multiple nearby pixels are averaged together, or via decimation, in which individual pixels are selected to represent larger blocks of pixels.  
The RGB values are further transformed into the YUV basis, where the luminance (Y) component may be used as a rough substitute for the RAW values when prioritizing frame rates is appropriate.

On-camera processing may then apply several corrections and transformations including lens-shading, which digitally corrects for the natural decrease in brightness as distance increases from the aperture. 

Each device is also configured to periodically collect geolocation data with which observation sessions and candidate events are labelled.
The operating system exposes an API which enables the software to request the highest-accuracy location information available, while limiting the frequency of power-intensive operations such as acquiring a Global Navigation Satellite System (GNSS) fix.
The system uses a combination of GNSS, WiFi, and cell tower information to determine location, typically within \SI{1}{m}.
On-board motion sensors are also used to determine whether a location update may be required.

The system time reported by some devices may exhibit significant drift.
When the device's operating system does not synchronize its clock with an external source sufficiently often, such as the Network Time Protocol (NTP) or Network Identity and Time Zone (NITZ) mechanism,  the system time can be highly inaccurate.
In order to standardize the accuracy of recording, the \crayfis~software includes a simple implementation of the NTP protocol~\cite{ntp}, which contacts the public servers at \verb|pool.ntp.org| to update the system time on a regular basis.

To-date, this structure and the associated algorithms below have been implemented only on the Android operating system as a proof-of-concept. An iPhone app utilizing the analogous iOS APIs is forthcoming.

\section{On-phone Corrections and Masking}

For the purposes of cosmic ray detection, the CMOS device should provide a uniform response, treating all pixels as equivalent independent detectors. In addition, the device should identify and remove defective pixels which provide false positives at a high rate, potentially dominating the limited upload bandwidth and overwhelming the CPU. 

\subsection{Lens-shading corrections}
Lens-shading is a standard image processing technique to compensate for the decreased intensity of light near the edges of a flat sensor, which are further from the aperture.
For the detection of cosmic rays, however, the optical lens and aperture is irrelevant; it is preferable that each pixel be treated as identical readout channels, so that a constant threshold can be applied to all pixels.
Often, these lens-shading gains are applied at the level of the sensor controller hardware or firmware and cannot be disabled via API controls.
In these cases, the lens shading must be reverse-engineered in order to quantify the spatially-dependent effective gains of each pixel, both as a systematic and as a component of an optimized trigger.

Though established methods exist for measuring per-pixel gains through controlled light exposures~\cite{belllabs,10.1117/12.175165}, these require a degree of user participation ill-suited for \crayfis. 
Instead, corrections must be computed exclusively from dark frames.

The lens-shading can be measured empirically if some model $X_i \sim p_i(x)$ of the underlying physical response of each pixel is known.
Since the lens shading factor $\lambda_i$ applied to each pixel is constant, it may be determined from the empirical average of the scaled value $Y_i = \lambda_i X_i$ via:
\begin{equation}
    \lambda_i = \left\langle Y_i\right\rangle / \left\langle X_i\right\rangle \,.
\end{equation}
However, in practice, the lens shading occurs before the final quantization of the digital pixel value; therefore, only the empirical average $\left\langle\lfloor\lambda_i X_i \rfloor\right\rangle$ is available.
For the low pixel means typical of dark frames, this does not provide a good approximation for $\left<\lambda_i X_i\right>$.

\begin{figure}[h]
    \centering
    \includegraphics[width=0.95\linewidth]{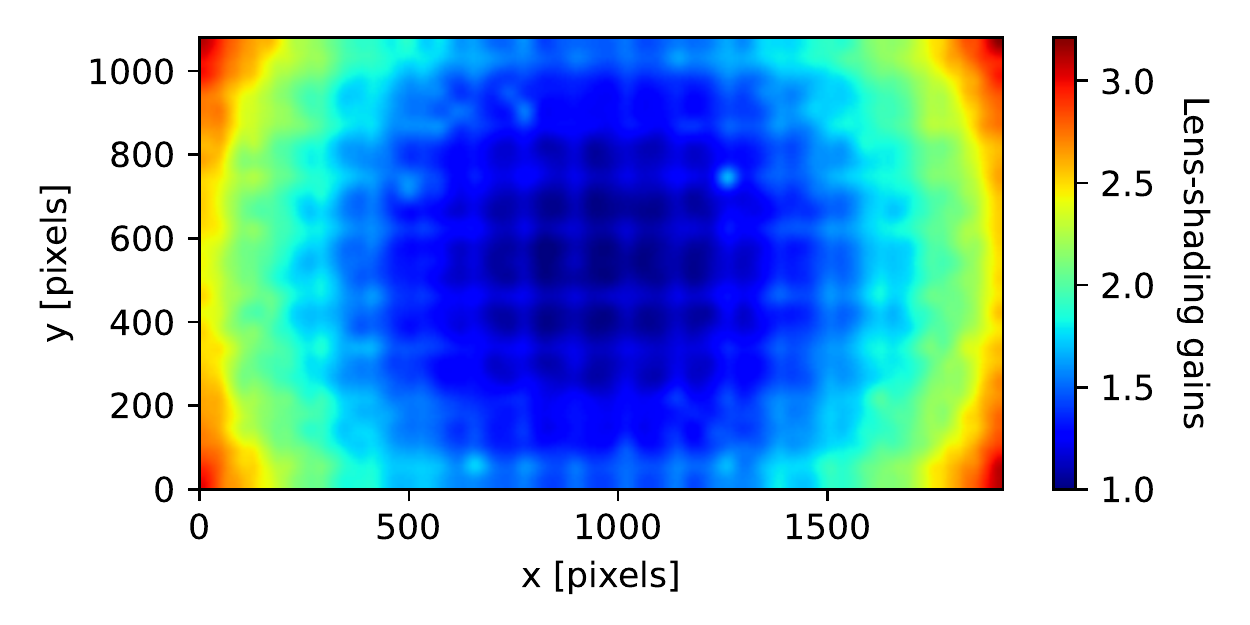}
    \caption{ Lens-shading gains across pixels of a Samsung Galaxy S7's CMOS sensor, estimated using Eq. \eqref{eq:scale_factor}. The observed aliasing pattern is a feature of the downsampling algorithm on the sensor rather than a defect in the calibration.}
    \label{fig:scale_hist_thresh}
\end{figure}

To enhance this approach, a functional form of $p_i(x)$ is assumed, from which the relationship between $\lambda_i \left\langle X_i\right\rangle$ and  $\left\langle\lfloor\lambda_i X_i\rfloor\right\rangle$ can be inferred. Denoting the probability density function of the undigitized dark current $\lambda_i X_i$ as $f_i(x)=p_i(x/\lambda_i)$, the mean of the truncated responses becomes:
\begin{equation}\label{eq:scale_factor_sum}
\left\langle\lfloor \lambda_i X_i\rfloor\right\rangle=\sum_{n=0}^\infty \int_n^{n+1}dx\,nf_i(x)=\sum_{n=1}^\infty \int_n^\infty dx\,f_i(x) \,.
\end{equation}
This is in general difficult to evaluate, and a numerical solution for several million pixels is computationally infeasible. However, with a convenient assumption that all pixels share a common distribution $p_i(x) = e^{-x/\mu}$, Eq. \eqref{eq:scale_factor_sum} can be simplified, yielding:
\begin{equation}
\left\langle\lfloor\lambda_i X\rfloor\right\rangle=\frac{1}{e^{\frac{1}{\lambda_i\mu}}-1}
\end{equation}
Inverted, this provides a closed-form correction for lens-shading: 
\begin{equation}\label{eq:scale_factor}
(\lambda_i\mu)^{-1} = \log(1+\left\langle\lfloor \lambda_i X \rfloor\right\rangle^{-1})
\end{equation}
where correction factors are normalized such that 
\begin{equation*}
\max_i(\lambda_i^{-1})=1\,.
\end{equation*}
This simple expression agrees reasonably well with numerical solutions of \eqref{eq:scale_factor_sum} involving more well-motivated noise models. Corrections computed for an example phone are shown in \Cref{fig:scale_hist_thresh}.

\subsection{Hot-pixel masking}

Each CMOS device contains defective pixels which provide false positives at a rate much higher than the typical pixel. Many such pixels trigger infrequently, but in aggregate either incur large CPU costs from an increased trigger rate or require more restrictive thresholds to keep the trigger rate reasonably low. Masking these pixels is therefore essential to optimizing device performance.

The basic strategy is to identify pixels with distributions of luminance values under dark conditions tending to larger values.  A rigorous statistical method such as a Kolmogorov-Smirnov test could accomplish this, but would require an impractically large sample size to distinguish more infrequent hot pixels. A simpler and more rapid approach is to examine the maximum value a pixel reports during a long dark run. However, some well-behaved pixels will record a single high value due to ionizing radiation; the second-highest value recorded by a pixel is then a simple but robust estimator of its performance, sufficient for an initial on-device masking.

\begin{figure}[h]
    \centering
    \includegraphics[width=0.9\linewidth]{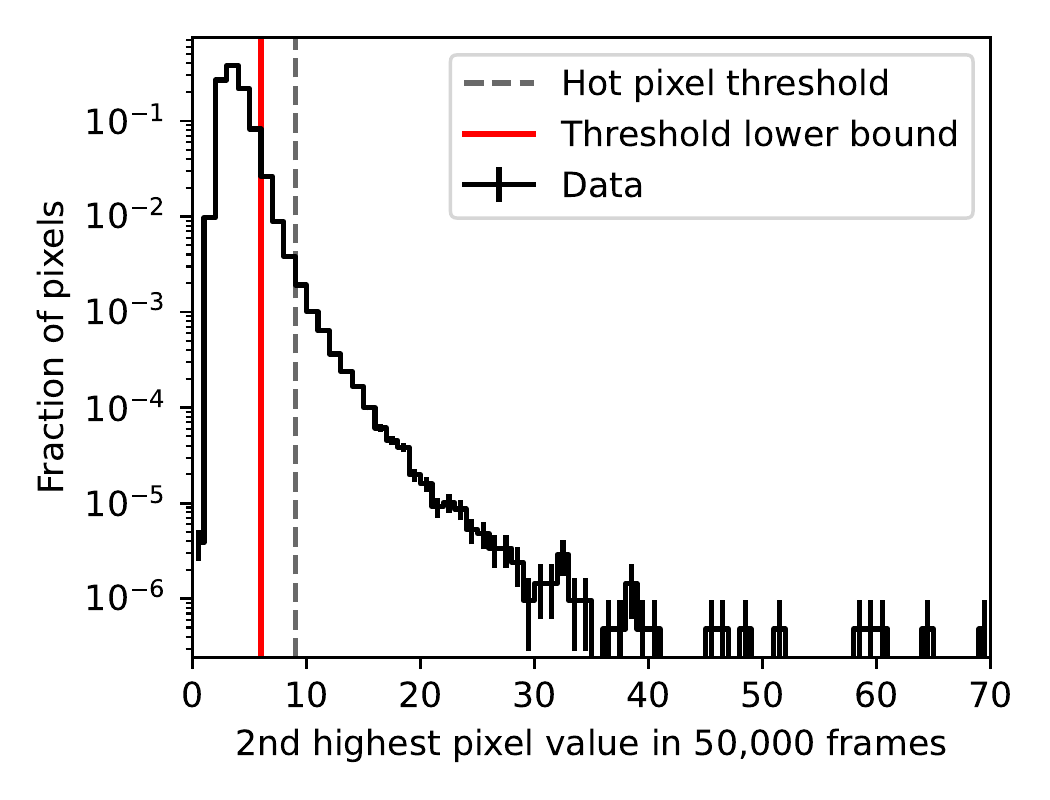}
    \caption{Distribution of the second-highest value recorded by each pixel in a dark run before correction for lens-shading, from which pixels are classified as clean or hot. The red line indicates the 99$^\mathrm{th}$ percentile of pixels; the threshold is always chosen to be at or above this level to prevent excessive masking.}
    \label{fig:second}
\end{figure}
\begin{figure}
    \centering
    \includegraphics[width=0.9\linewidth]{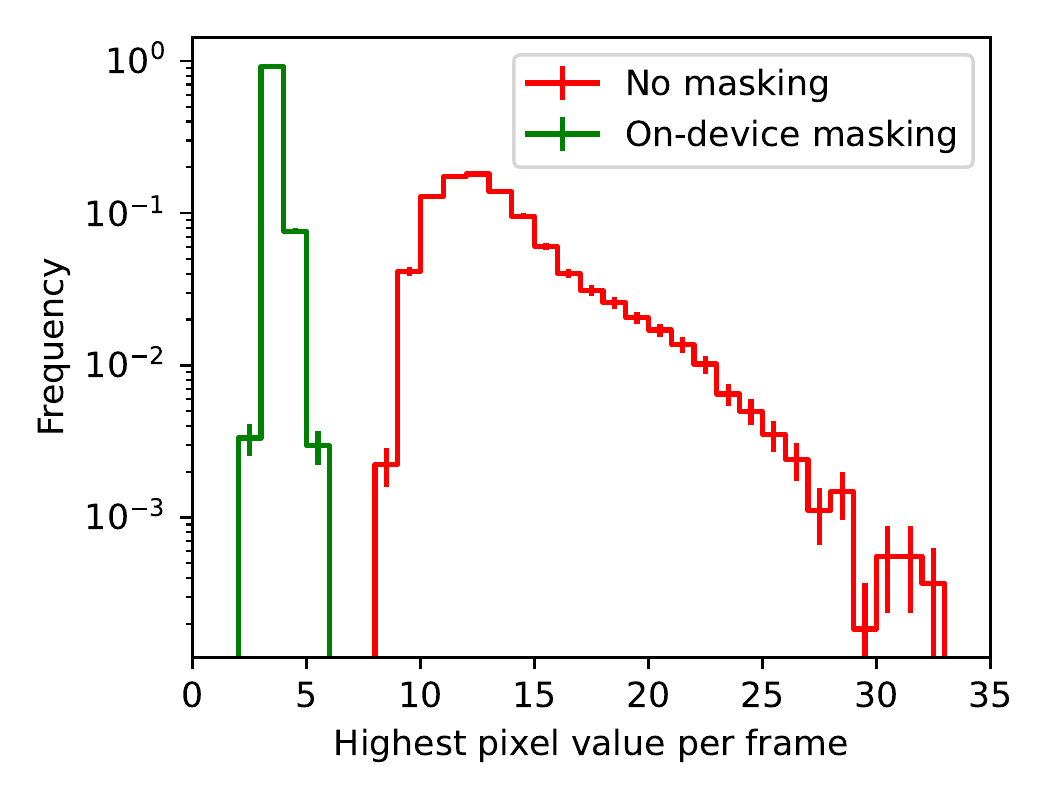}
    \caption{Demonstration of the effects of hot-cell masking. Shown is the highest pixel value per frame with and without on-device hot-cell masking.}
    \label{fig:hotcell}
\end{figure}

Pixels are classified as hot if their second-highest value in a long dark run falls above a threshold, selected such that pixels above threshold share their second-highest pixel value with fewer than 0.2\% of pixels; see \Cref{fig:second}.  To prevent excessive masking, the threshold is not allowed to sink below a value which would remove more than 1\% of pixels. Typical performance is demonstrated in \Cref{fig:hotcell}. More exhaustive hot-pixel masking requires substantially larger sample sizes and is left to the off-device system due to constraints from the device RAM.

\subsection{Data}

The lens-shading correction and hot-pixel masking affect the efficiency of the CMOS response to incident particles, and the chosen parameters must be recorded for later analysis of the data. 
Several features minimize the storage requirements of these parameters.
By downsampling the grid of lens-shading corrections, re-normalizing the values as 8-bit integers, and compressing it as a JPEG, the total size is typically reduced to less than a kilobyte with no more than a 0.5\% loss in precision. 
To then keep the metadata consistent between the phones and the server, the app operates on the same compressed JPEG buffer received by the server.
Updates to these parameters from off-device software, discussed below, are stored as incremental changes.
Lens-shading and hot-pixel parameters are cached on both the server and the phone, such that storage-intensive on-device parameter selection is in practice never repeated.

\section{Trigger and Operation}

\subsection{Trigger threshold}

If computing power, storage, and data bandwidth were unlimited, the experiment would prefer to record and transmit every pixel in every frame. For more realistic operation, the primary function of the on-device software is to scan the calibrated clean pixels and identify the small fraction of excited pixels which are the best cosmic ray candidates. In a captured frame which contains a candidate, most of the frame is empty, and saving and transmitting such a sparse image in its entirety is inefficient and unnecessary.  

To identify and transmit high quality candidates, a two-level trigger system is used. The first level examines each frame, rejecting those which have no clean pixels above a fixed threshold. The second level examines each pixel, storing those which have luminance above a second threshold and their neighbors. The pixel-level threshold can be  decoupled from the frame threshold, allowing greater resolution of faint tracks without the cost of higher frame pass rates.

The trigger thresholds are chosen by the on-device software to achieve a remotely-configurable frame pass rate during an initial calibration stage, described below.
\Cref{fig:rates} shows the frame and pixel pass rates as a function of trigger thresholds. Decreasing the thresholds may increase the efficiency within a given frame, but would create a burden on the CPU which prevents running at higher resolutions and frame rates. A default target rate of \SI{0.33}{Hz} is employed to balance these priorities. Such thresholds have been associated with muon and photon spectra in Refs. \cite{crayfis2015,crayfis-muons}, albeit with different devices and sensor resolutions.

\subsection{Operation}

The on-device software operates as a state machine with seven states. Five of these---\textsc{Init}, \textsc{Survey}, \textsc{Precalibration}, \textsc{Calibration}, and \textsc{Data}---sequentially structure the calibrations and DAQ. The last two---\textsc{Idle} and \textsc{Finished}---handle breaks in data-acquisition due to insufficient battery, overheating, or interruptions from the user.

The \textsc{Init} state first activates, inspects, and configures the device hardware necessary for \crayfis~to operate via the Android APIs.
The camera sensor is configured with as many image preprocessing effects disabled as the device will allow.
The ISO gain is set at the maximum analog value supported by the camera, allowing the greatest precision when specifying pixel thresholds in terms of photo-electron counts.
In order to minimize memory turnover and to utilize the onboard GPU, the sensor output buffer is managed with Android's Renderscript API, which can perform certain array operations quickly and efficiently.
The location services are configured to receive high-accuracy updates from a combination of GNSS, cell tower, and WiFi signals, though at infrequent intervals to conserve power.
Several other sensors are also utilized: the barometer is employed as a secondary proxy for altitude, and the accelerometer and magnetic field sensors are used to jointly reconstruct the phone's orientation. 

With these configured, the \textsc{Survey} state then attempts to locate a covered camera sensor suitable for data-taking. If the sensors indicate that the phone is not lying flat, or if substantial light levels are still found on all image sensors, the user will be prompted to reposition the phone; if this fails after several attempts (such as when the phone is being actively used apart from the \crayfis~UI) and the phone is actively charging, the app will sleep in the \textsc{Idle} state and reattempt several minutes later; otherwise, the app will exit. This assures that the \crayfis~app does not interfere with the user’s phone use and does not unnecessarily discharge the battery, both of which are essential to a positive user experience. These checks on the phone orientation and light levels are continued for all subsequent frames, prompting an intermission in the \textsc{Idle} state if the camera is no longer properly covered. The phone orientation is nonetheless recorded to monitor small shifts in the phone's position, a possible source of light pollution.

\begin{figure}
    \centering
    \includegraphics[width=0.9\linewidth]{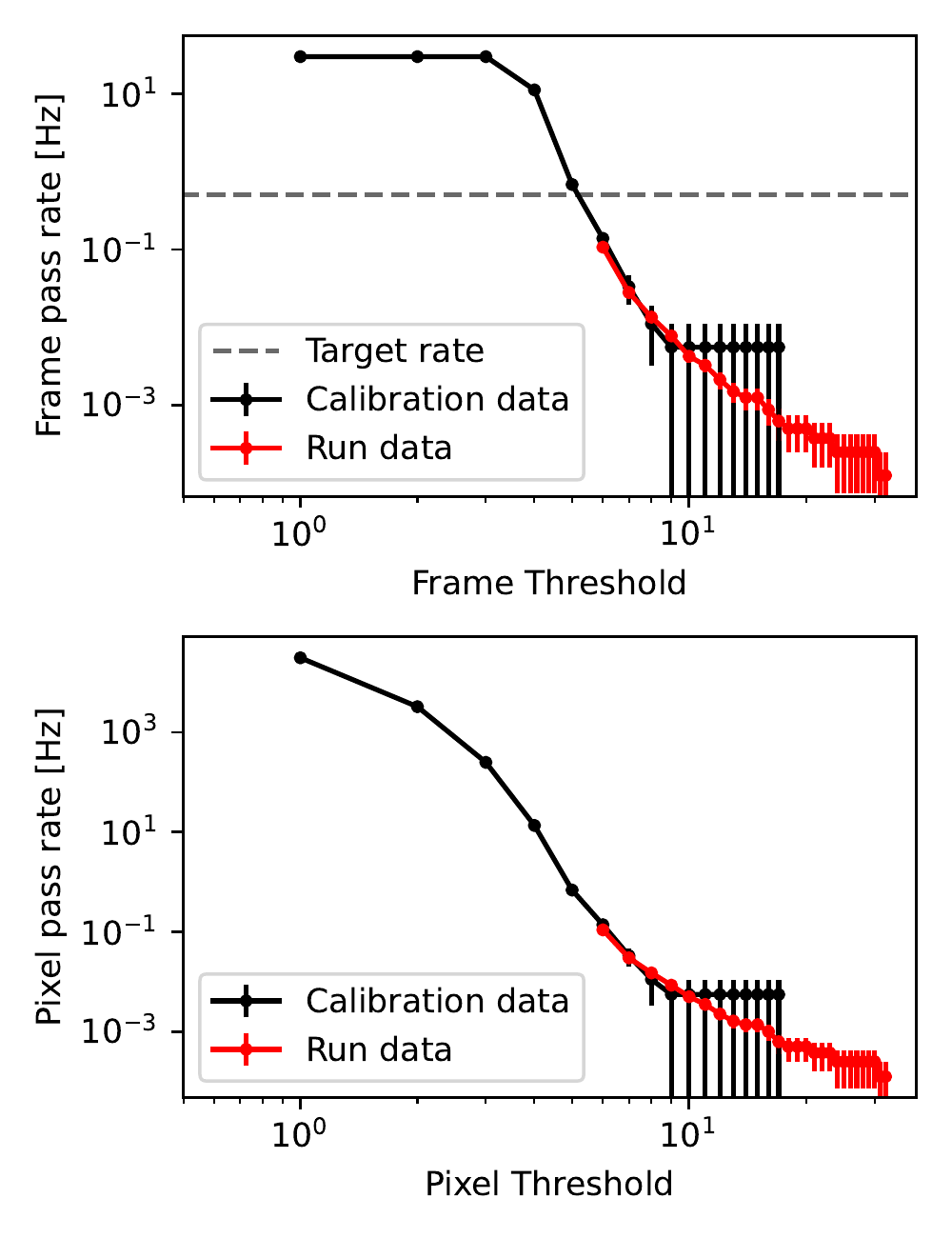}
    \caption{Rate of frame triggers (top) and resulting pixel triggers (bottom) as a function of threshold on a sample phone. The default \SI{0.33}{Hz} target rate of frame triggers is also drawn above. For simplicity, the bottom plot assumes equal thresholds for the frame and pixel triggers.}
    \label{fig:rates}
\end{figure}

With an appropriate camera to stream buffers, the \textsc{Precalibration} state then determines the lens-shading and hot-pixel parameters, either by loading cached data from the server or phone, or through an on-device measurements when no such cache is available. The \textsc{Calibration} state then selects trigger thresholds as shown in \Cref{fig:rates}.

With thresholds chosen, triggering of cosmic-ray candidates can begin in the \textsc{Data} state. When a frame and pixel pass the respective thresholds, its coordinates and value are saved; the pixel values of surrounding 5×5 blocks are also recorded, though in such a way when multiple pixels are triggered, no pixel value is saved twice. A separate zero bias trigger periodically captures random 10×10 blocks of pixels, from which noise variations at large and small scales can be quantified offline.

\subsection{Performance optimization}

To reduce the impact on the device's resources, the on-device software is factorized into separate background (DAQ) and foreground (UI) components. Accordingly, the device can enter sleep mode, eliminating the CPU burden of foreground processes, while data acquisition continues. To assure that the DAQ does not run indefinitely and consume the user's battery as a background process, frames are screened for a flat camera orientation, and multiple consecutive fails stop the DAQ if the UI is inactive.

Frame processing is likewise made far more efficient by employing the GPU through Android's Renderscript APIs. By design, the GPU is optimized for asynchronous operations across large arrays, and is able to perform the same trigger operations far more efficiently than the standard Java layer. In particular, Renderscript features a highly-optimized method for building histograms of pixel values in frame buffers, which reduces processing time in the frame trigger by two orders of magnitude compared to a Java implementation. However, Renderscript is specifically designed for asynchronous, element-wise computation such as array-to-array mappings. Several frame-processing operations during calibration fall outside this domain, for which native OpenCV~\cite{opencv} is employed instead.

\begin{figure}
    \centering
    \includegraphics[width=0.9\linewidth]{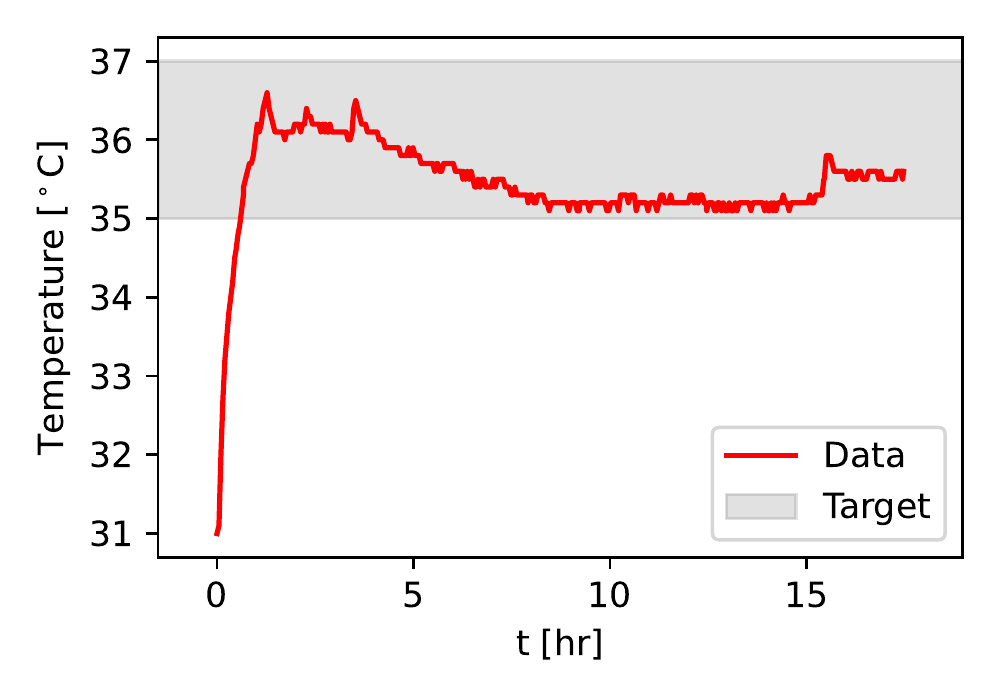}
    \caption{Battery temperature versus time during typical running on a sample phone at 1080p and 30 FPS. The target range of 35--37℃, used for server-side resolution and frame rate adjustments, is shaded.}
    \label{fig:perf}
\end{figure}

To quantify the burden on the CPU, battery temperature is periodically queried. As the resources allocated to other applications may vary on a nightly basis, this metric is more useful for scaling the \crayfis~app in real time than its absolute CPU usage. A number of phones in the 2012--2014 generation, which exceed 45℃ at 1080p and 30 FPS without these modifications, instead reach a plateau at 35--37℃; nonetheless, temperatures above 41℃ induce an intermission in the \textsc{Idle} state as a safeguard. \Cref{fig:perf} shows the temperature of a sample phone under typical running conditions.
When the temperature is consistently below the target range, the phone's resolution or frame rate may be increased to improve data quality.

\subsection{Data format}

The on-device software collects several pieces of information to provide context for triggered pixels, called an “\verb|Event|.”  
Every three minutes, these are assembled into an \verb|ExposureBlock|, a set of frames with identical exposure and trigger settings. Each block contains the following data:

\begin{itemize}[label=, leftmargin=0in]
\item \textbf{ExposureBlock}
\begin{itemize}[label=--, leftmargin=0.15in]
  \item Unique device ID
  \item Run ID
  \item Start/end time
  \item DAQ State: \textsc{(Pre)Calibration}/\textsc{Data}
  \item Initial location: latitude, longitude, altitude, precision
  \item Sensor resolution
  \item Trigger settings
  \item Number of streaming frames not processed (e.g. due to CPU latency)
  \item Histogram of all pixel values scanned
  \item \textbf{Event}(s)
  \begin{itemize}[label=--, leftmargin=0.15in]
    \item Frame time
    \item Location: latitude, longitude, altitude, precision
    \item Phone orientation vector (x, y, z)
    \item Histogram of all pixel values (triggered and untriggered) in frame
    \item \textbf{Pixel}(s)
    \begin{itemize}[label=--, leftmargin=0.15in]
      \item Position (x, y) of triggered pixels
      \item 8-bit values of surrounding 5x5 block.
    \end{itemize}
  \end{itemize}
\end{itemize}
\end{itemize}

\noindent The Run ID associates each \verb|ExposureBlock| with another data structure containing hardware and software metadata such as the particular camera in use.

When the cache of \verb|ExposureBlock|s exceeds 50 kB, these are serialized using Google Protocol Buffers and uploaded to the server; with the default trigger rates described above, this corresponds to a network load of 50--100 Bps per active device. A cryptographic hash is used to verify the data received by the server, computed with a salt appended to the uploaded request body. 
This can be used to prevent malicious data from polluting the experiment: without the salt, an attacker is unable to create passable data that the server will accept. 

\section{Off-device software}

\begin{figure}[h]
    \centering
    \includegraphics[trim=50 0 50 0, clip, width=\linewidth]{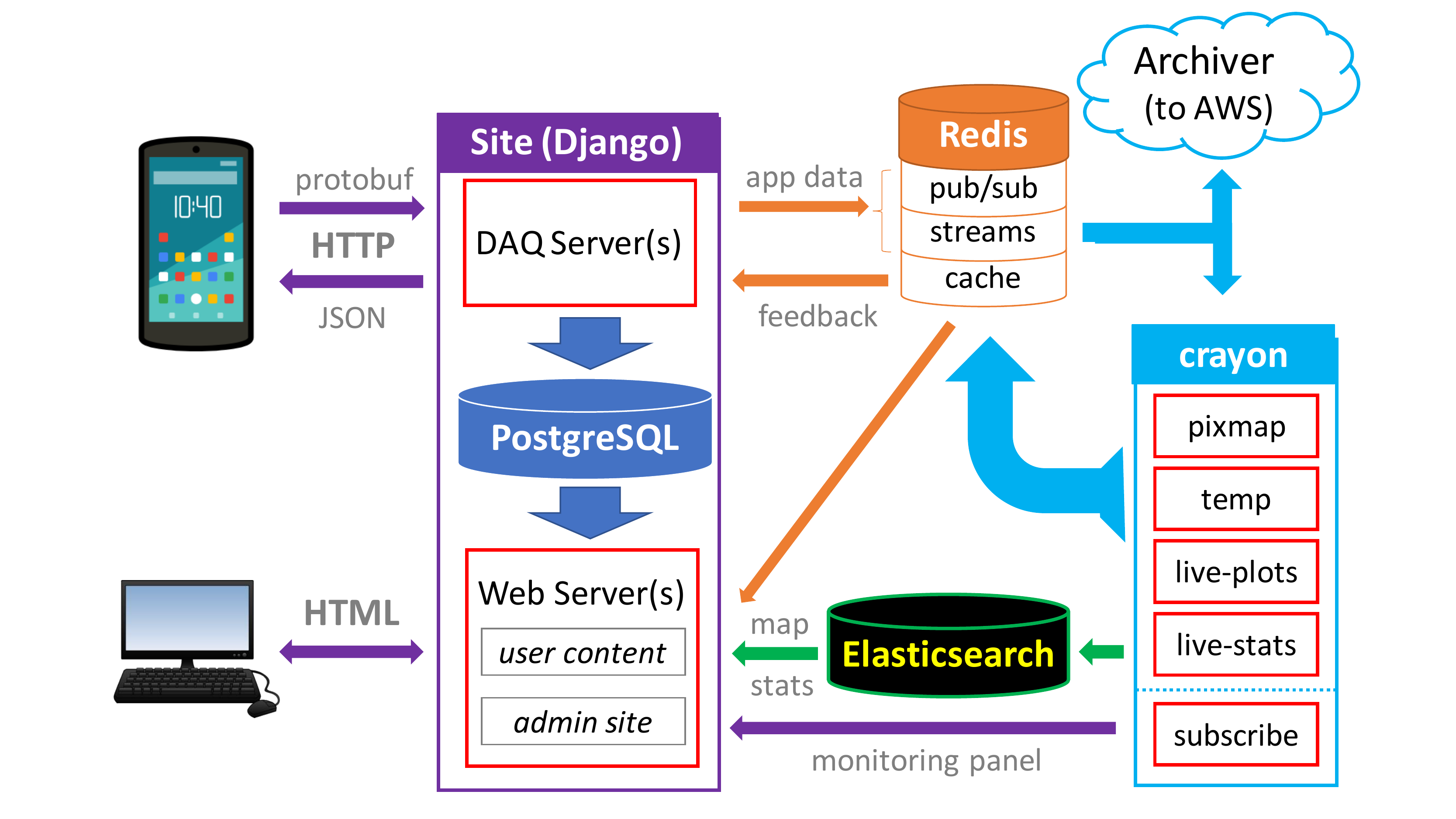}
    \caption{Organization and interaction of the elements of the off-device software.}
    \label{fig:diagram_webapp}
\end{figure}

The off-device software is responsible for receiving data from the smartphone, cataloguing it for later scientific applications, providing a web-based user interface to monitor and organize device operation, and refining calibration for individual devices. 

The software operates in the cloud, and relies on a server to provide the web interface,  three databases (Redis, PostgreSQL, and Elasticsearch), several “\verb|crayon|” (CRAYfis ONline) worker nodes to analyze data on the fly  and the “Archiver,” a node routing data beyond the web app into offline data storage; see \Cref{fig:diagram_webapp}. 
These components are containerized through Docker and run in a Kubernetes ecosystem~\cite{k8s}, a self-healing and self-scaling cluster environment.

The PostgreSQL database stores structured experimental metadata, linking each run to a particular user. 
It can efficiently process simple queries, such as the statistics populating each user's personal page. 
The Elasticsearch database, meanwhile, handles much of the same data, but is optimized for sophisticated queries such as locating sets of devices which are geographically clustered. 
The Redis database stores unstructured data, primarily uploaded application data which is streamed to the Archiver as well as to the five \verb|crayon| workers for online processing.

Two \verb|crayon| workers, \verb|crayon-pixmap| and \verb|crayon-temp|, provide  feedback loops  which compute adjustments to particular devices and cache the results in Redis. 
Two more, \verb|crayon-live-plots| and \verb|crayon-live-stats|, are responsible for updating figures on the website. 
Finally, \verb|crayon-subscribe|  creates detailed live-monitoring  for devices of interest.

\subsection{Operating parameter refinements}

Devices can derive the parameters of lens-shading corrections and hot-pixel masking themselves, but the greater scalability and memory afforded by cloud-based computing permits more robust calculations of operating parameters, which are then provided to individual devices. Each device first attempts to obtain parameters from the server. If this fails, it will use its own most recent copy. As a fallback, the device will derive lens-shading and hot-cell masking parameters itself.

A worker node, shown in \Cref{fig:diagram_webapp} as ``\verb|crayon-pixmap|," maintains histograms of the spatial coordinates of triggered pixels, from which device operating parameters are adjusted. 

For hot-pixel masking, a sparse matrix format provides the necessary single-pixel precision; however, as more pixels are triggered over months of data-taking in a device, the sparse format becomes increasingly memory-intensive. To avoid wasting memory on pixels with normal trigger rates, a rolling average is maintained when the number of \verb|ExposureBlock|s exceeds 250. 
If $k$ new \verb|ExposureBlock|s with trigger occupancies $n(x,y)$ are added to the existing sparse matrix $N_t(x,y)$, the updated matrix $N_{t+1}(x,y)$ is given by:
\begin{equation}
N_{t+1}(x,y)=\frac{250}{250+k}(N_t(x,y)+n(x,y)) \,.
\end{equation}
Entries in the matrix are purged when their value decreases below $0.01$ to limit the number of nonzero entries. 

Pixels with rates above a certain threshold are masked; currently, this stands at 15 triggers per 250 \verb|ExposureBlock|s, or 1 per 60,000 frames. 
We require 15 triggers to limit our rate of false positives, as this masking is permanent. 

The downsampled grid of lens-shading parameters, denoted as $w(x,y)=\lambda(x,y)^{-1}$, is similarly adjusted by \verb|crayon-pixmap|. 
The spatial distribution of pixel rates is monitored, binned to achieve the same dimensions as this downsampled grid. 
If the variance of this distribution exceeds the mean by a threshold, (currently $\sigma^2/\mu > 1.8$), a correction is calculated and applied. 
The mean Poisson rate across the entire sensor is used to calculate the $p$-value for each bin $p(x,y)$ via the Poisson CDF. 
An inverse logistic response function provides the correction:
\begin{equation}
\Delta w(x,y)=\frac{\alpha}{\sqrt{N}}\log(p(x,y)^{-1}-1)
\end{equation}
where $\alpha$ sets the rate of convergence and $N$ is the total number of triggers in the histogram. 
This function has the desirable property that corrections approach positive and negative infinity for $p$-values of 0 and 1, respectively; to prevent overcorrection, $|\Delta w(x,y)|$ is limited to $0.125$. 
Furthermore, as the response becomes more uniform, larger amounts of data will be required to pass the threshold for a correction, and thus the $N^{-1/2}$ dependence will lead to smaller adjustments.

\begin{figure}
    \centering
    \includegraphics[width=0.9\linewidth]{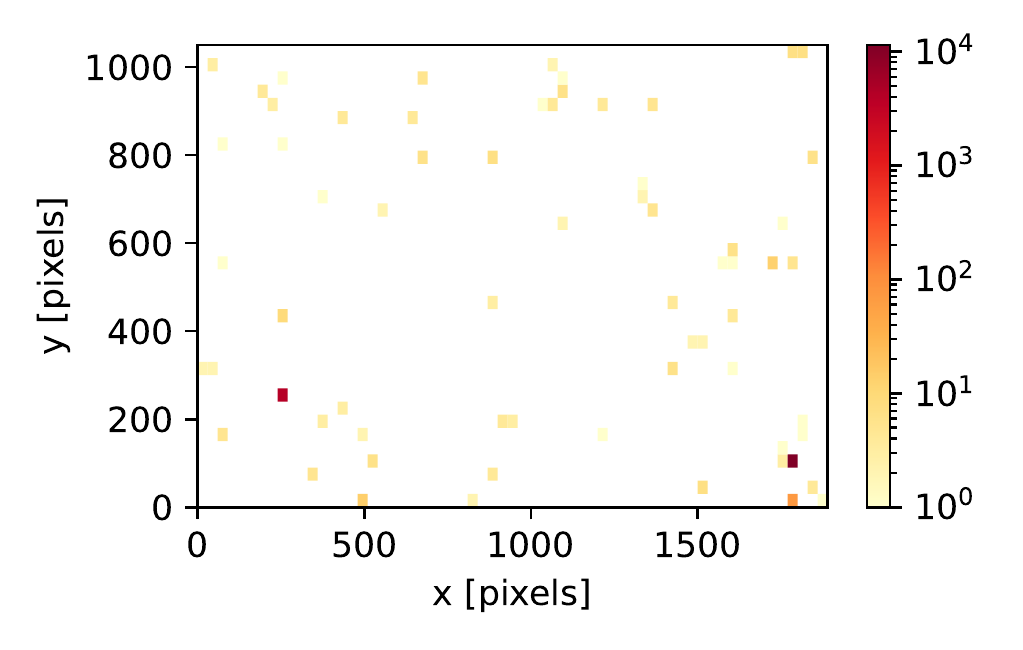}
    \includegraphics[width=0.9\linewidth]{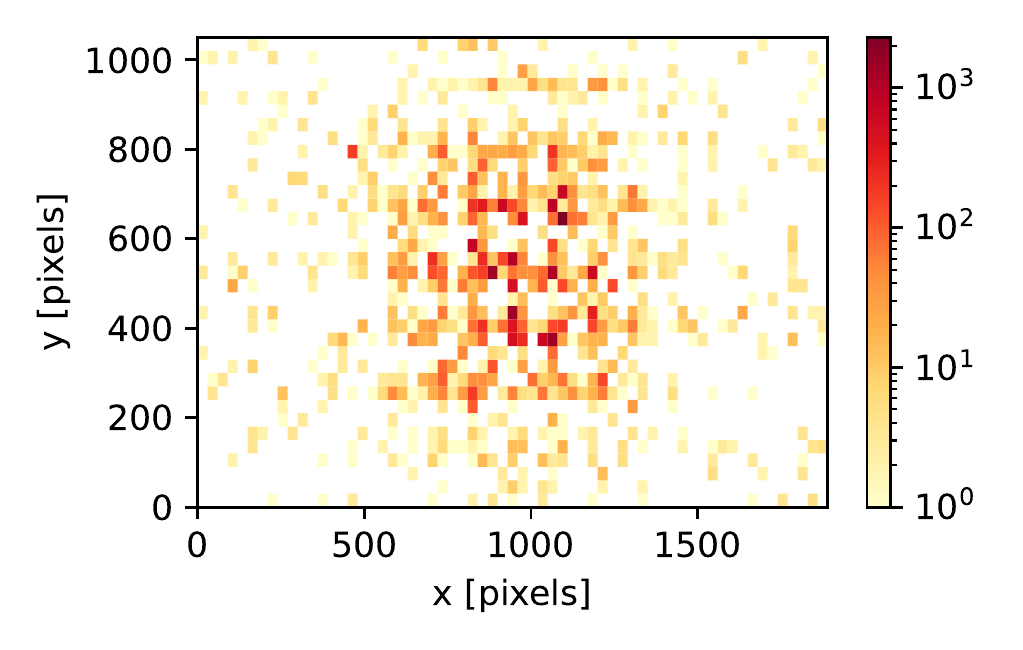}
    \includegraphics[width=0.9\linewidth]{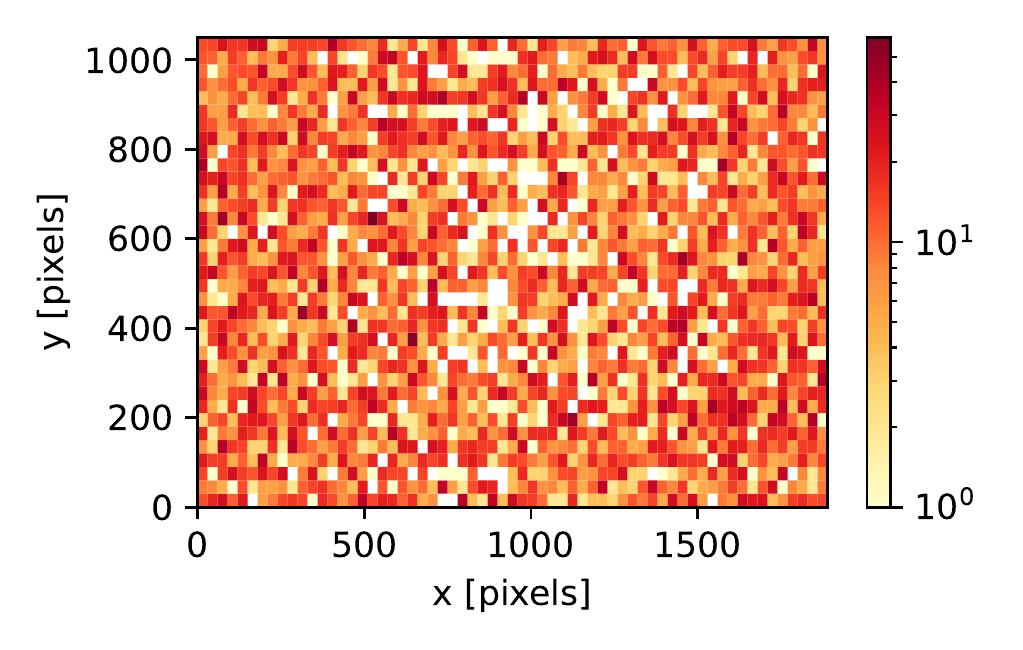}
    \caption{ Frequency of pixel hits by location in a Samsung Galaxy S7 under three stages of hot pixel masking and lens shading corrections. The top panel is before any lens-shading or hot-pixel masking, in which the sensor's response is dominated by a few hot pixels. The center panels is after initial on-device lens-shading and hot-pixel masking, which improves the uniformity but leaves a significant population of less-active hot pixels in the center. The bottom panel uses refined off-device parameters calculated after seven days of dark running.}
    \label{fig:pixmap_results}
\end{figure}

\Cref{fig:pixmap_results} shows the stages of this process for a device streaming at 1080p. The positions of pixels which pass the triggers are shown before any lens-shading or hot-pixel masking, after initial on-device lens-shading and hot-pixel masking, and after application of refined parameters calculated  off-device after seven days of running. The on-device lens-shading and hot-pixel masking are clearly an improvement, but slightly over-correct for lens-shading, and leave a number of less-active hot pixels near the center of the sensor. This is a consequence of the ordering of the calibrations, as hot pixel masking precedes correcting for lens-shading gains. A second round of hot-pixel masking with scaled pixel values would greatly reduce this hot pixel population, but in practice, the thresholds are already sufficiently reduced (\Cref{fig:hotcell}) to commence data-taking, leaving additional hot-pixel masking for the server. The final lens-shading corrections and hot-pixel masking show a nearly uniform sensor response. 

\subsection{Thermal management}

Ideally, the \crayfis~on-device software would sample frames at the highest resolution and frame rate possible.
However, while the CMOS hardware is generally capable of very high throughput sampling, the memory bandwidth and computational power of the phone becomes a limiting factor.
Even when limiting throughput to the highest rate that is computationally feasible, the smartphone's CPU and battery will inevitably heat up to temperatures which can further limit performance as well as degrade the hardware's lifespan.
The on-device software is designed to monitor hardware thermometry, from which off-device feedback mechanisms manage the CPU load and maintain a safe operating temperature through automatic adjustments to the frame rate, resolution, and trigger thresholds.
In future updates, the relative priority of these adjustments will be customized to specific device models, both automatically and manually.

\section{Experiment Management}

\begin{figure*}
    \centering
    \includegraphics[width=\linewidth]{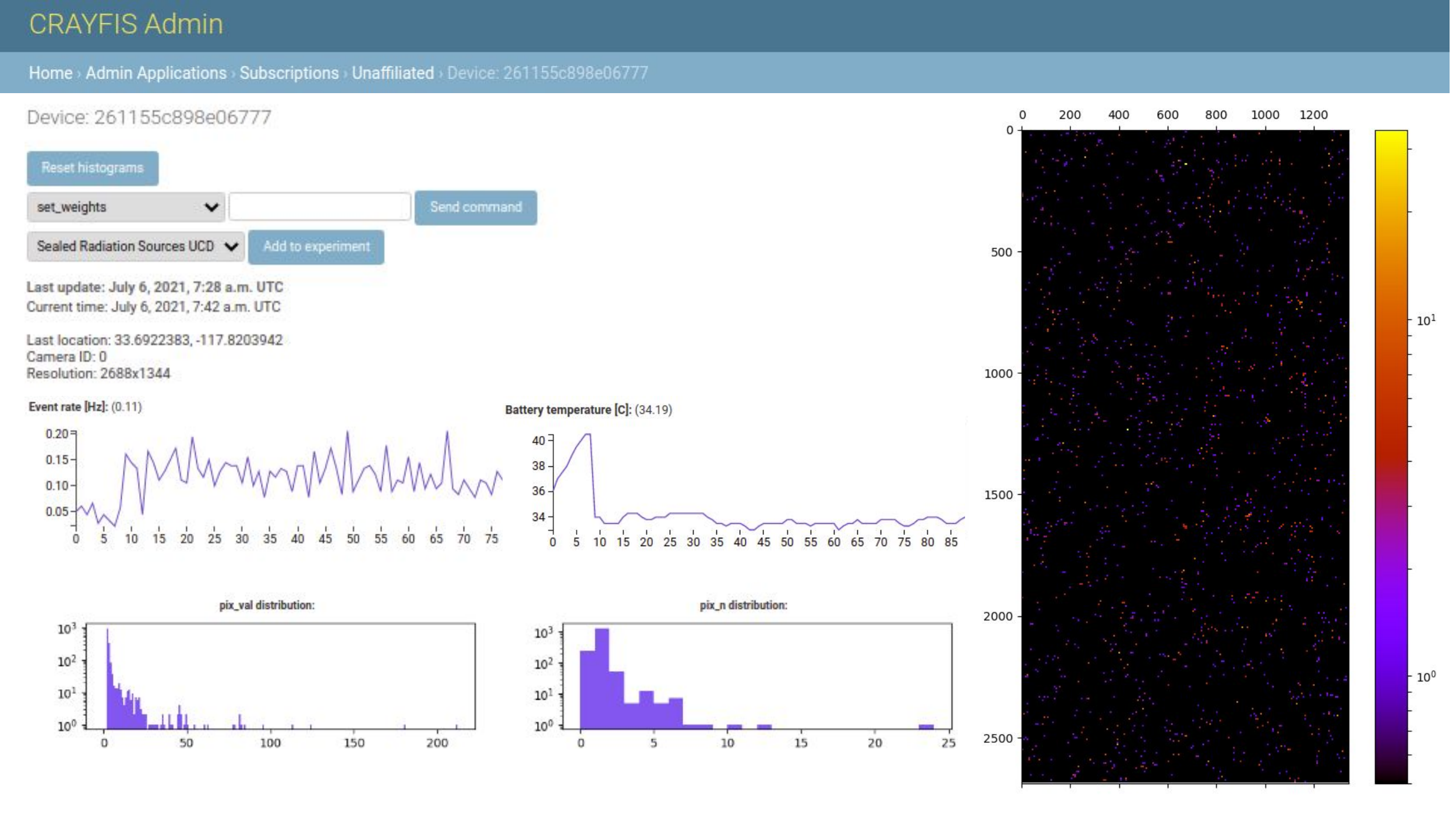}
    \caption{A partial view of the monitoring panel for a particular device. Recorded metrics include average pixel values, battery temperature, and thresholds versus time, as well as histograms of pixel values and hit coordinates.}
    \label{fig:management}
\end{figure*}

A monitoring application is powered by the “\verb|crayon-subscribe|” worker, allowing experimenters to subscribe to specific devices. Incoming data from such devices are cached in Redis, from which detailed graphs of performance metrics can be generated, shown below in \Cref{fig:management}.

This monitoring interface is linked with Elasticsearch, allowing both mass subscriptions with targeted queries and mass feedback to certain device classes. This is implemented through an “ElasticCommand” application, which allows, for example, all phones of a specific model or region, or a set of phones exhibiting unusual behavior, to be collectively monitored and manipulated remotely. This serves as an effective way to optimize and standardize performance. For instance, when the lens-shading pattern of a particular phone model is well-known through controlled laboratory testing, the computed scale factors can be directly applied to all phones of that model, bypassing the standard online calibration entirely. This principle can also be extended to optimizing frame rates, resolutions, trigger rates, ISO gain, and even the camera buffer settings particular to a specific model. 

In addition, the server-side control provides crucial experimental controls over the operational parameters of the hardware on the network.
This allows experimenters to design specific run conditions, e.g. acquiring data with higher rates and lower thresholds for a limited time, or experiments based on geo-fencing, where all devices within a defined geographic region can be configured in a particular way.

\begin{figure}
    \centering
    \includegraphics[trim=240 50 220 50, clip, width=\linewidth]{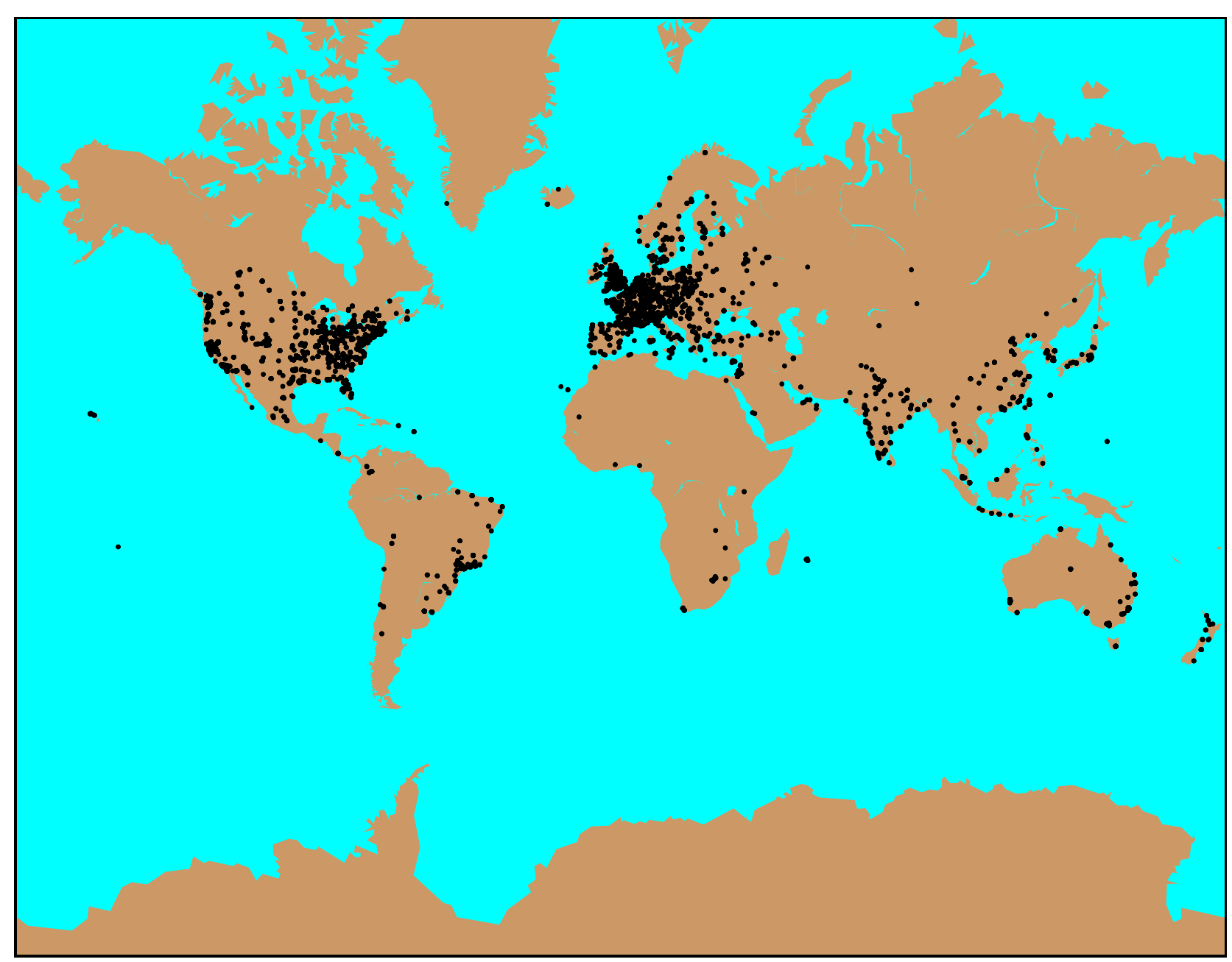}
    \caption{Global distribution of the $\sim 25\,000$ devices which have uploaded data to the \crayfis~servers.}
    \label{fig:global}
\end{figure}

\section{Conclusions}

Repurposing the global network of smartphones as a scientific instrument capable of generating data using on-board instruments presents enormous opportunities as well as challenges. We have described a scalable, flexible, global data acquisition system which has an initial globally-distributed base of users; see Fig~\ref{fig:global}.  The system includes on-phone components to perform remote calibration and data collection while minimizing impact to the owner, as well as off-phone components which manage the network and are capable of storing the data for later scientific analysis and refining the operating parameters of individual phones.   
\\[\baselineskip]

\section{Acknowledgements}

JS was supported by a generous grant from the Jenkins Family Foundation.  The authors are grateful to Andrew Nelson, Jodi Goddard, Homer Strong, Jay Karimi, Kyle Cranmer, and Kyle Brodie for their contributions in software development, and to Serguei Kolos and Gokhan Unel for helpful comments on earlier drafts.

\bibliography{prod}

\clearpage
\clearpage

\end{document}